\documentclass{moriond}
\usepackage{amsmath}

\def\be{\begin{equation}}
\def\ee{\end{equation}}
\def\bea{\begin{eqnarray}}
\def\eea{\end{eqnarray}}

\newcommand{\Photo}{\includegraphics[height=35mm]{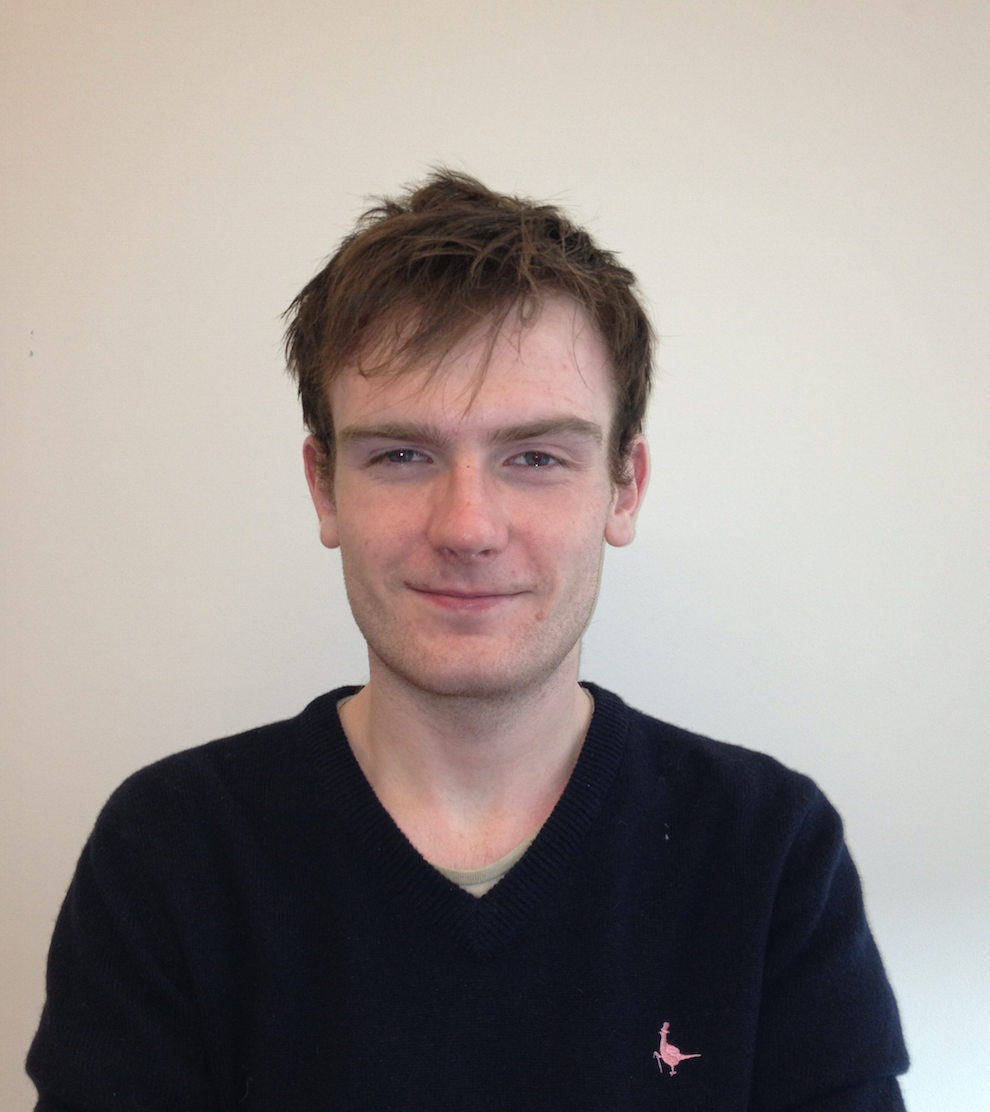}}

\begin{document}
\vspace*{4cm}
\title{Searching for Sterile Neutrinos with MINOS}

\author{ A. Timmons \\ on behalf of the MINOS collaboration }

\address{School of Physics and Astronomy, University of Manchester,\\
Manchester M13 9PL, England}

\maketitle\abstracts{
This document presents the latest results for a 3+1 sterile neutrino search using the $10.56 \times 10^{20}$ protons-on-target data set taken from 2005 - 2012. By searching for oscillations driven by a large mass splitting, MINOS is sensitive to the existence of sterile neutrinos through any energy dependent deviations using a charged current sample, as well as looking at any relative deficit between neutral current events between the far and near detectors. This document will discuss the novel analysis that enabled a search for sterile neutrinos setting a limit in the previously unexplored regions in the parameter space $\{\Delta m^{2}_{41}, \sin^2\theta_{24}\}$. The results presented can be compared to the parameter space suggested by LSND and MiniBooNE and complements other previous experimental searches for sterile neutrinos in the electron neutrino appearance channel.}

\section{Introduction}
Since the initial significant observation of neutrino oscillations in 1998, 2001~\cite{ref:SuperK,ref:SNO} and 2002 for anti-neutrinos~\cite{ref:KamLAND}, physicists across the world have designed multiple generations of experiments designed to confirm and probe the nature of neutrino oscillations using accelerator neutrinos~\cite{ref:OPERA,ref:T2K,ref:K2K,ref:Nova}, solar neutrinos~\cite{ref:SAGE,ref:GALLEX}, and nuclear reactor antineutrinos~\cite{ref:Chooz,ref:Reno,ref:DoubleChooz,ref:DayaBay}. 


The majority of neutrino oscillation experiments have obtained model-independent evidence for neutrino oscillations that are compatible with the three-flavour model; a model with three flavour eigenstates $\left(\nu_{e}, \nu_{\mu}, \nu_{\tau}\right)$ and three mass eigenstates $\left(\nu_{1}, \nu_{2}, \nu_{3}\right)$ that mix described by the $3 \times 3$ PMNS rotation matrix~\cite{ref:PMNS}. Neutrino oscillations are energy dependent and are governed by the difference of the square of the mass eigenstates, $\Delta m^{2}_{32}$ and $\Delta m^{2}_{21}$, while the amount of mixing (the amplitude) is governed by three mixing angles $\theta_{12}, \theta_{13}, \theta_{23}$, and a $CP$ violating phase $\delta_{13}$. Recent global values of these parameters are presented by the Particle Physics Data group (PDG)~\cite{ref:PDG}. 

There have been several anomalous results within the neutrino community that, among other explanations, can be explained by the existence of sterile neutrinos. Two examples are from the Liquid Scintillator Neutrino Detector (LSND) and MiniBooNE short-baseline experiments, which observed an excess of $\overline{\nu}_{e}$ that are incompatible with the three-flavour model~\cite{ref:LSND,ref:MiniBooNE}. A potential explantation is for a large mass splitting of the order $\Delta m^{2} \sim \mathcal{O} \left(1\, \mbox{eV}^{2}\right)$, this however, is in disagreement with the two globally measured mass splittings from the three-flavour model. To accommodate for this extra mass splitting the three-flavour model has to be extended to allow for additional neutrino flavour and mass eigenstates; these additional neutrinos, in order to agree with the LEP measurements of the Z-boson decay rate and line shape~\cite{ref:LEPZdecay}, would not interact via the weak interaction and are hence called \emph{sterile}.

\section{The MINOS Experiment}

The MINOS experiment has two steel-scintillator calorimeters~\cite{ref:MINOSNIM} designed to be functionally equivalent. Both detectors are made of alternating layers of $1.00~\text{cm}$ thick plastic-scintillator and $2.54~\text{cm}$ thick steel planes. The neutrino beam is provided by the Neutrinos at the Main Injector (NuMI) neutrino beam~\cite{ref:NUMI,ref:NUMI2} based at Fermilab. As neutrinos travel through the detector they interact with the iron-nuclei, and charged final-state particles travel through the scintillator depositing energy which is read out as light via wavelength shifting fibres and photomultiplier tubes. The steel planes are magnetised by a coil aligned to the longitudinal axes of each detector. The trajectories of the charged particles are therefore curved allowing MINOS to distinguish $\nu_{\mu}$ and $\overline{\nu}_{\mu}$ charged current (CC) interactions within the detectors. 

The Near Detector (ND) is situated 1.04~km downstream from the neutrino target at Fermilab. With a mass of 0.98~kton the ND measures the reconstructed neutrino energy spectrum, in a three-flavour model this is before oscillations have occurred. 

The Far Detector is 735~km downstream from the neutrino production target, 705~m underground in a mineshaft in northern Minnesota. The FD is significantly larger than the ND to compensate for the decrease in the neutrino flux. With a mass of 5.4~kton the FD measures the reconstructed neutrino energy spectrum and will observe a different neutrino flavour composition of the beam due to neutrino oscillations.

\section{The MINOS 3+1 Sterile Neutrino Analysis}

The MINOS experiment was originally built for the measurement of the three-flavour atmospheric oscillation parameters $\theta_{23}$ and $\Delta m^{2}_{32}$ by looking at $\nu_{\mu}$ disappearance looking at CC events with an $L/E$ optimised at $500$~km/GeV at the FD. This analysis considers the 3+1 sterile neutrino phenomenological model which requires a $4 \times 4$ PMNS matrix. This gives rise to an extra mass splitting $\Delta m^{2}_{41}$, three additional mixing angles $\theta_{14}$, $\theta_{24}$, $\theta_{34}$ and two additional CP violating phases $\delta_{14}$, and $\delta_{24}$. This analysis will use two channels, the first is the muon survival probability $P(\nu_{\mu} \rightarrow \nu_{\mu})$, by looking at muon neutrino disappearance, any deviations from the three-flavour oscillation probability would occur due to mixing with sterile neutrinos states. The 3+1 muon neutrino survival probability can be approximated to:

\begin{equation}
 P(\nu_{\mu} \rightarrow \nu_{\mu} ) \approx 1 - \sin^{2}2 \theta_{23}\cos 2 \theta_{24} \sin^{2} \Delta_{31} - \sin^{2}2\theta_{24}\sin^{2}\Delta_{41},
 \label{eqn:muonapprox}
\end{equation}

\noindent giving sensitivity to $\sin^2\theta_{24}$. The second channel is through sterile neutrino appearance $P(\nu_{\mu} \rightarrow \nu_s)$ which can be approximated to:

\begin{equation}
1 - P(\nu_{\mu} \rightarrow \nu_{s} ) \approx 1 - c^{4}_{14}c^{2}_{34}\sin^{2}2\theta_{24}\sin^{2}\Delta_{41} - A\sin^{2}\Delta_{31} - B\sin^{2}2\Delta_{31}. 
 \label{eqn:NCapprox}
\end{equation}

The expression $1 - P(\nu_{\mu} \rightarrow \nu_{s} )$ can be thought of as disappearance of neutral current (NC) events (identical to sterile neutrino appearance). The terms $A$ and $B$ are functions of the mixing angles and phases. To first order $A = s^{2}_{34}\sin^{2}2\theta_{23}$ and $B = \frac{1}{2}\sin\delta_{24}s_{24}\sin2\theta_{34}\sin2\theta_{23}$. Equation~\ref{eqn:NCapprox} shows dependence on the sterile parameters $\theta_{24}$, $\theta_{34}$ and $\delta_{24}$. The sensitivity to $\delta_{24}$ is limited by the poor resolution (due to the outgoing neutrino) and significant background from $\nu_{\mu}$ and $\nu_{e}$ CC events, therefore the assumption is made that $\delta_{13} = \delta_{14} = \delta_{24} = 0$ along with assuming CPT conservation. The sterile mixing angle $\theta_{14}$ does not appear in equation~\ref{eqn:muonapprox} due to being a sub-dominant term. It appears in equation~\ref{eqn:NCapprox}, in both cases $\theta_{14}$ only becomes non-negligible for large values, an analysis of solar and reactor neutrino data yields the constraint $\sin^{2}\theta_{14} < 0.041$ at 90\% C.L.~\cite{ref:th14limit} therefore this analysis sets $\theta_{14}$ = 0, which is interpreted as no mixing between $\nu_{e}$ and $\nu_{s}$ i.e $|U_{e4}|^{2} = 0$.

\subsection{Event Selection}
This analysis uses two different data samples, a sample of CC-$\nu_{\mu}$ ($\nu_{\mu} N \rightarrow \mu X$) events and a sample of NC ($\nu X \rightarrow \nu X'$) events. 

A sample of CC-$\nu_{\mu}$ is selected by searching for events with a $\mu$ track in the final state with possible hadronic activity. To distinguish CC events from NC events, four topological variables are studied using a $k$-Nearest-Neighbour algorithm~\cite{ref:Rus}. The four variables used are: the number of MINOS detector planes associated with a muon track (muon tracks tend to extend much further than NC showers), the average energy deposited per scintillator plane along the track, the transverse energy deposition profile, and the variation of the energy deposited along the muon track. At the ND the CC sample has an efficiency of 53.9\% and purity of 98.7\%, at the FD the efficiency is 84.6\% and purity of 99.1\%. The low efficiency at the ND is due to the removal of events with $\mu$ tracks that end in and around the magnetic coil hole due to poorly understood data/MC discrepancies.

A sample of NC events is selected by searching for events with activity that spread fewer than 47 steel-scintillator in the detector. Events with a track require that the track go no further than five planes beyond the hadronic shower. Additional cuts are required to remove poorly reconstructed events that may containment the sample due to multiple coincident events that could potentially confuse the event reconstruction software. At the ND the NC sample has an efficiency of 79.9\% and purity of 58.9\%, at the FD the efficiency is 87.6\% and purity of 61.3\%. The biggest background comes from high elasticity CC-$\nu_{\mu}$ events. 

\subsection{Systematic Uncertainties}
The systematic uncertainties used in this analysis are incorporated into the fit using a covariance matrix $V$. The covariance matrix contains both statistical and systematic uncertainties on the Far over Near reconstructed neutrino energy ratio (F/N),

\begin{equation}
 V = V^{\text{stat}} + \sum\limits_{i=1}^N V^{\text{syst}}_{i}.
\end{equation}

The statistical uncertainty is less than 24\% in each energy bin with an average of 15\%. The dominant systematics will be discussed in this text below,

\begin{itemize}
\item{The relative normalisation between the FD and ND for the CC sample is 1.6~\% and 2.3~\% for the NC sample.}
\item{The uncertainty on the acceptance and selection efficiency of the ND is evaluated by varying event selection requirements in data and MC to probe weaknesses in the simulation. The discrepancies between data/MC are taken as the uncertainty on the F/N neutrino energy spectrum. This systematic is energy dependent and includes correlations between energy bins and varies between $2$~\% and $6$~\% for the CC sample and is below $0.6$\% in the NC sample.}
\item{The procedure used to reduce the number of poorly reconstructed events from the NC sample uses a variable that is not perfectly modelled by simulation. The mis-modelling between data/MC is used to produce an uncertainty for the NC sample which is energy dependent and includes bin to bin correlations; the size of this systematic falls from 5\% below 1~GeV to less than 1.5\% above 5~GeV. }
\item{The rest of the evaluated systematics take into account hadron production, beam focusing, neutrino cross-section and uncertainty on the CC contamination in the NC sample and the NC contamination in the CC sample. The total uncertainty on the F/N energy spectrum arising from these less dominant systematic sources sums (in quadrature) to no more than 4\% in any parts of the energy spectra.}
\end{itemize}

\begin{figure}[!ht]
	\centering
	\includegraphics[height=0.30\textwidth]{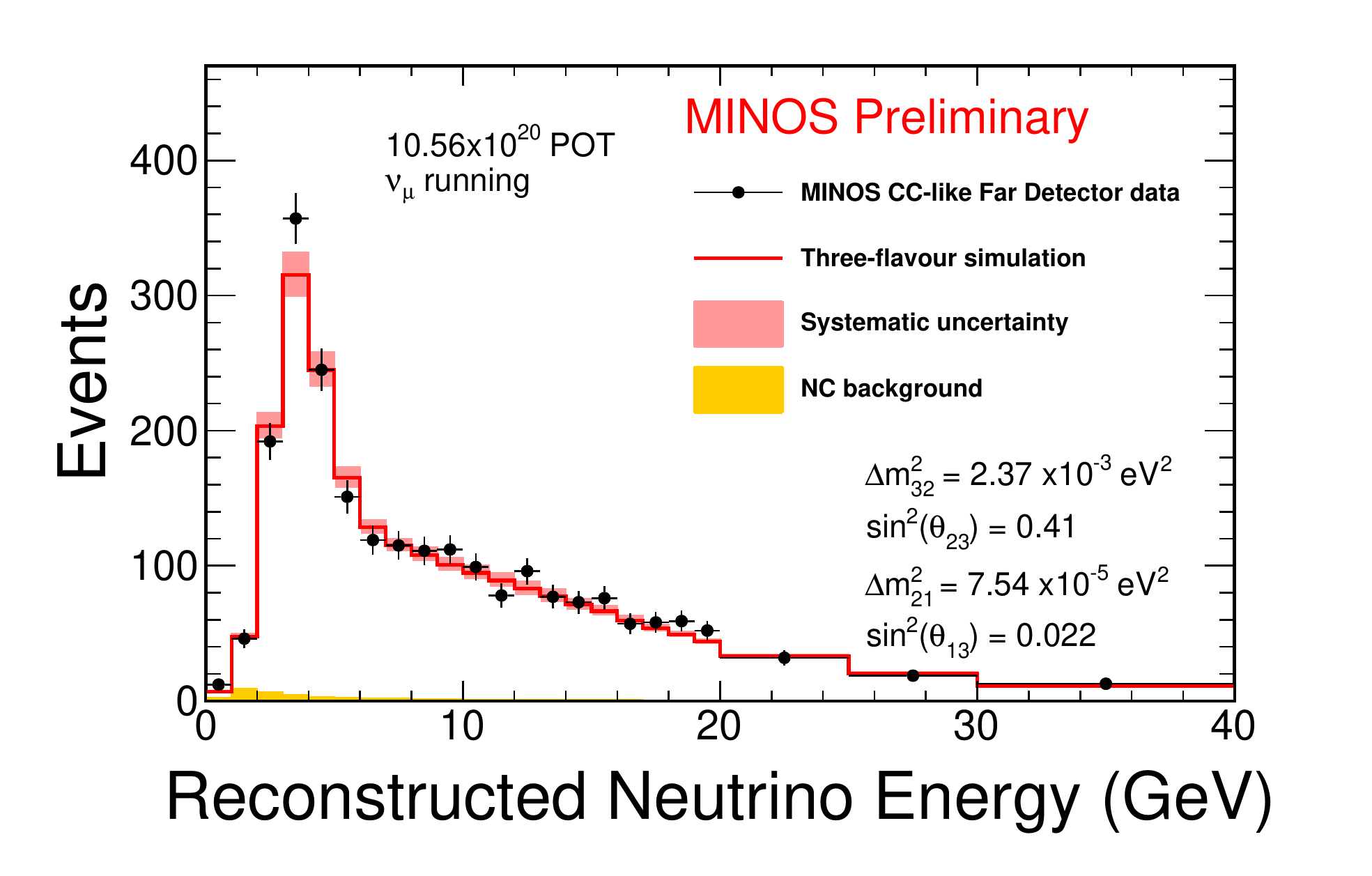}
	\includegraphics[height=0.30\textwidth]{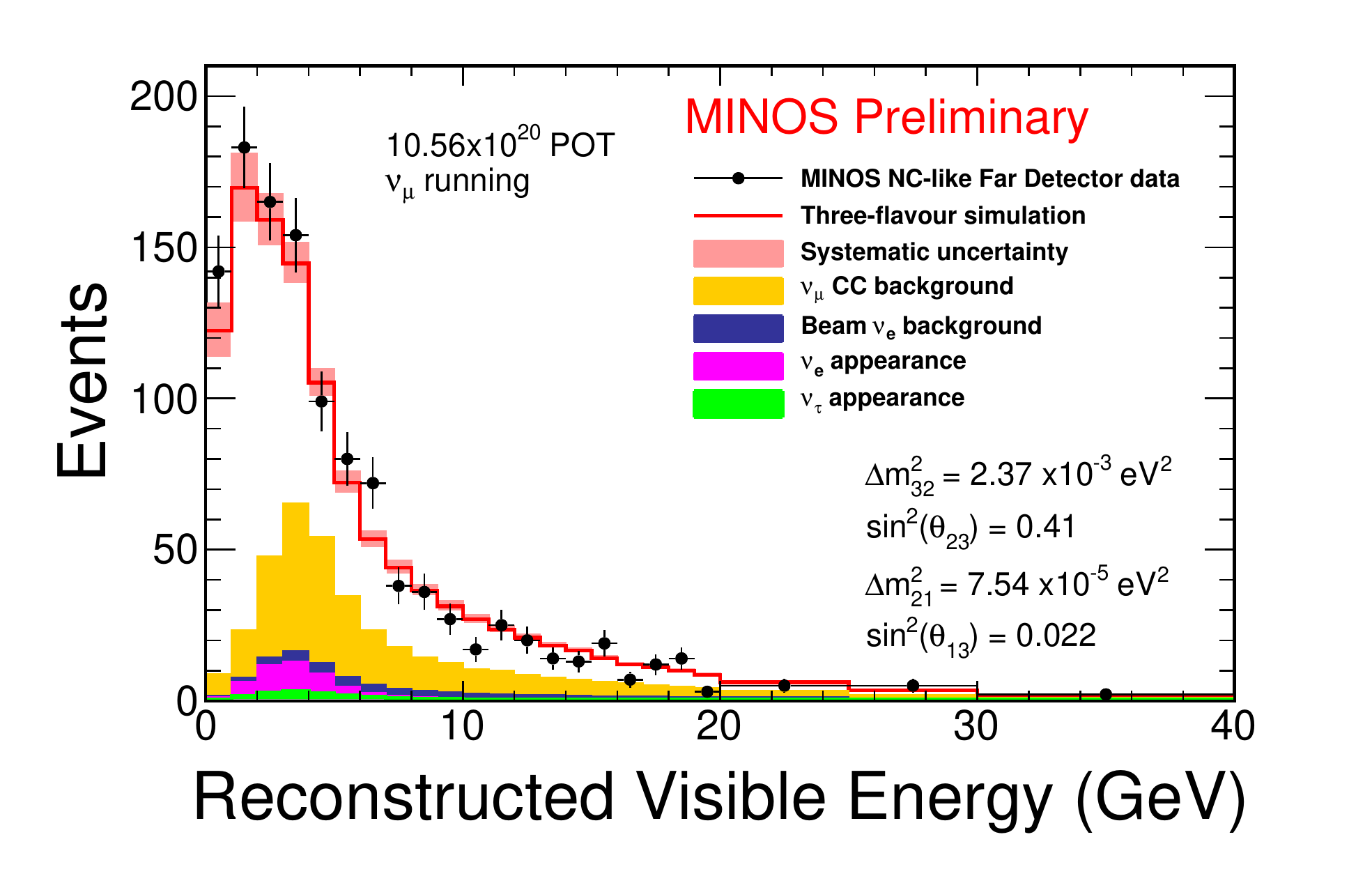}
	\caption{The CC (left) and NC (right) reconstructed neutrino energy spectrum using the MINOS dataset at the FD. The red histograms in both panels are the three-flavour expected neutrino energy spectrum for the extrapolated FD prediction with systematic error band. The values used for the three-flavour prediction were not fitted to this spectrum but taken from the latest MINOS three-flavour analysis~\protect\cite{ref:threeflavPRL}. In both cases the relevant backgrounds have been calculated for CC and NC and are included.}
	\label{fig:FDspectrum}
\end{figure}

\subsection{A model independent approach}

Looking for a deficit of NC events from a three-flavour neutrino model does not require an assumption for how many additional sterile neutrinos are present and can be seen as model-independent. A FD three-flavour prediction is generated by weighting the raw FD simulation in each energy bin by the discrepancies from the ND data/MC comparison. The CC and NC FD prediction for both data and three-flavour prediction can be seen in figure~\ref{fig:FDspectrum}; no large deviations from the three-flavour model can be seen. 

To better quantify the agreement between expectation and observation at the FD NC energy spectrum, a metric $R$ is defined as:

\begin{equation}
	R = \frac{N_{\text{data}} - \sum B_{\text{NC}}}{S_{\text{NC}}},
\end{equation}

\noindent where $N_{\text{data}}$ is the integrated number of NC events from the FD prediction in figure~\ref{fig:FDspectrum} in the energy range from $0-40\,\,\text{GeV}$.  The integrated total NC background is given as $\sum B_{\text{NC}}$ using the truth information from simulation, with $S_{\text{NC}}$ defined as the integrated sum of true NC events. The $R$ value is calculated over three different energy ranges motivated by the shape of the NC neutrino spectrum having a large portion events at low energies. Table~\ref{tab:Rvalues} shows the calculated $R$ values with statistical and systematic uncertainties. A value of $R = 1$ coincides with complete agreement between data and the three-flavour predicted FD neutrino energy spectra.

\begin{table}[ht!]
	\centering 
	\caption{R values calculated from the FD NC neutrino energy spectrum.}
	\vspace{0.3cm}
	\begin{tabular}{c |c |c |c |c}
        \hline
	Energy (GeV) & R value & $\pm$ syst & $\pm$ stats & $\pm$ total  \\ [0.5ex] 
	\hline  
         0 - 40              & 1.049    & 0.095         & 0.045           & 0.105  \\ [1ex]
         0 - 3                & 1.100    & 0.073         & 0.061           & 0.095  \\ [1ex]
         3 - 40              & 1.008    & 0.128         & 0.067           & 0.144 \\ [1ex]
         \hline
	\end{tabular}
	\label{tab:Rvalues} 
	\end{table}

\subsection{Performing a fit using the 3+1 sterile neutrino model}

\begin{figure}[!ht]
	\centering
	\includegraphics[height=0.60\textwidth]{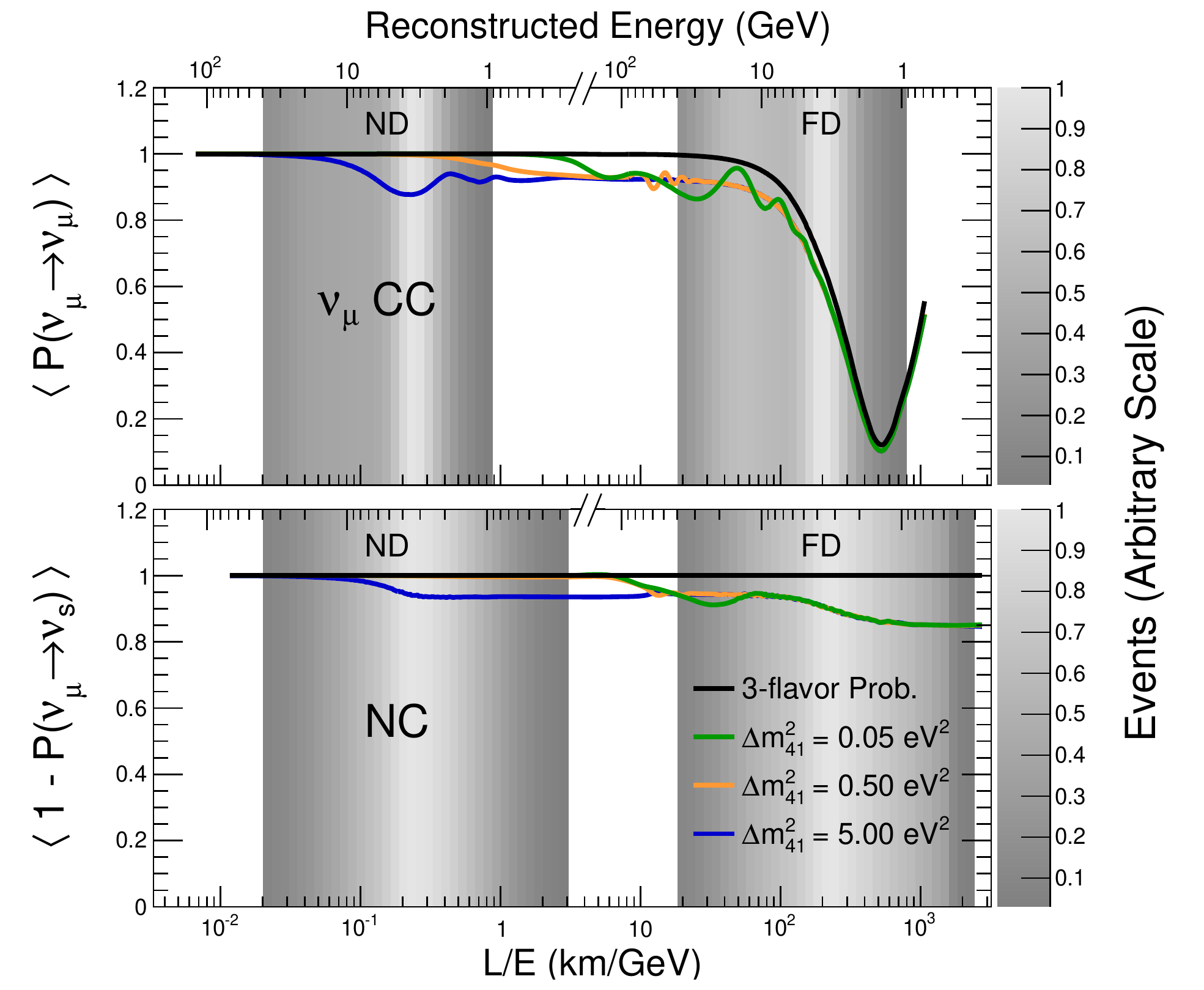}
	\caption{The $\nu_{\mu}$-CC and NC disappearance probabilities as a function of $L/E$ for various values of $\Delta m^{2}_{41}$. The figure illustrates how for large values of $\Delta m^{2}_{41}$ that the traditional FD extrapolation techniques used by MINOS in previous analyses will no longer work. Values of $\Delta m^{2}_{21}$, $\Delta m^{2}_{32}$, $\theta_{23}$, $\theta_{13}$ and $\theta_{12}$ taken from the PDG~\protect\cite{ref:PDG}, sterile parameters used are $\theta_{24} = 0.15$, $\theta_{14} = 0.2$, $\theta_{34}$ = 0.5 and all CP phases set to zero.}
	\label{fig:LoEstacked}
\end{figure}

MINOS analyses have traditionally used predicted FD energy spectra as a function of the oscillation parameters constrained by ND data. However, this analysis considers a 3+1 model where the additional mass splitting $\Delta m^{2}_{41}$ is unknown and the range of interest for MINOS is $10^{-4} - 10^{2}~\text{eV}^{2}$. Once $\Delta m^{2}_{41} > 1$eV$^{2} $, the neutrino oscillation probability becomes non-zero in the region of $L/E$ probed by the ND, thus both $\nu_{\mu}$ and $\overline{\nu}_{\mu}$ disappearance and sterile neutrino appearance occurs at the ND; figure~\ref{fig:LoEstacked} shows the oscillation probability for muon (anti)neutrino disappearance and sterile neutrino appearance for increasing values of $\Delta m^{2}_{41}$. Non-zero oscillation probabilities at the ND makes the extrapolation method using ND data to create a FD prediction invalid. 

\begin{figure}[!ht]
\centering
\includegraphics[height=0.30\textwidth]{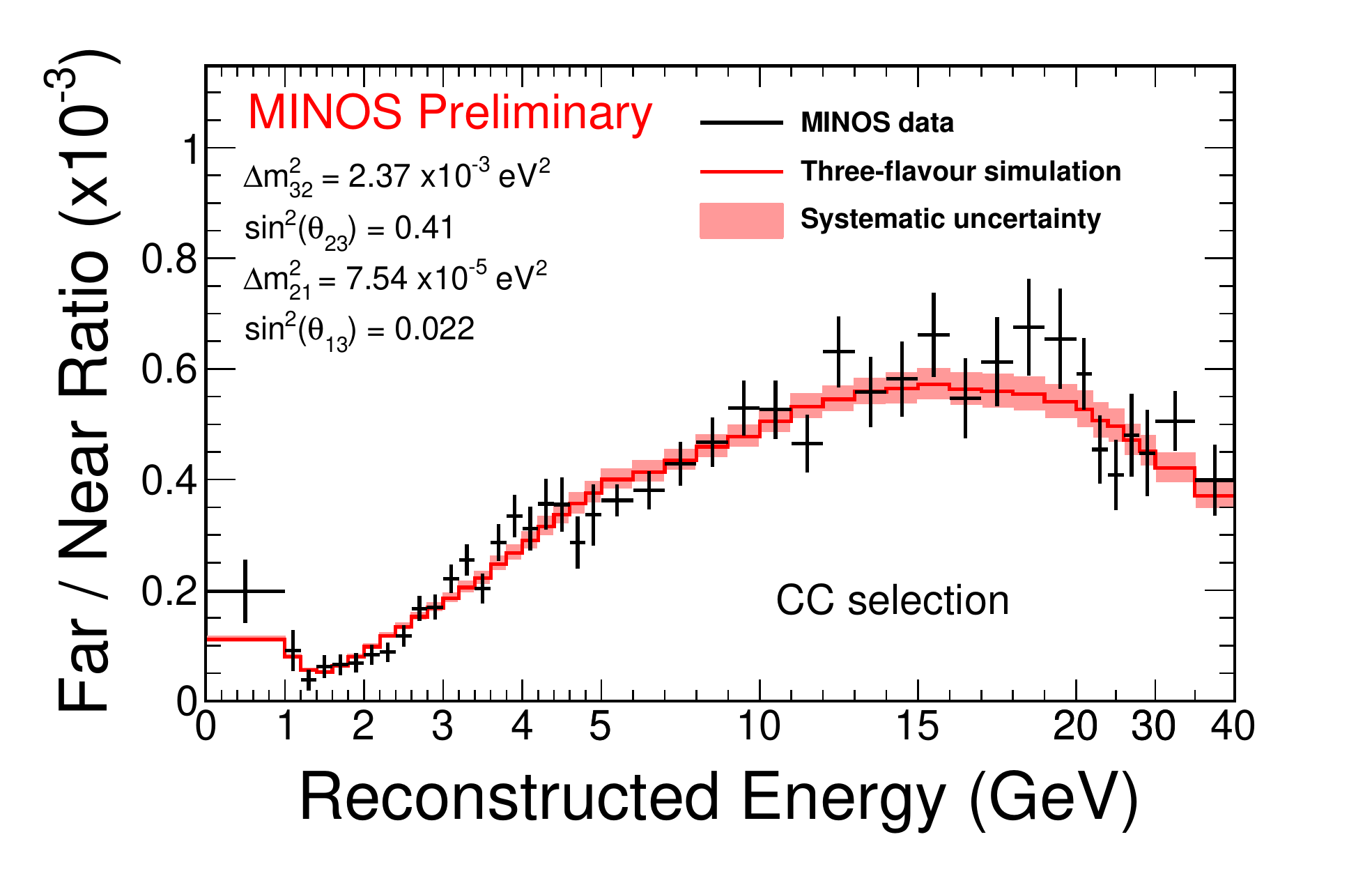}
\includegraphics[height=0.30\textwidth]{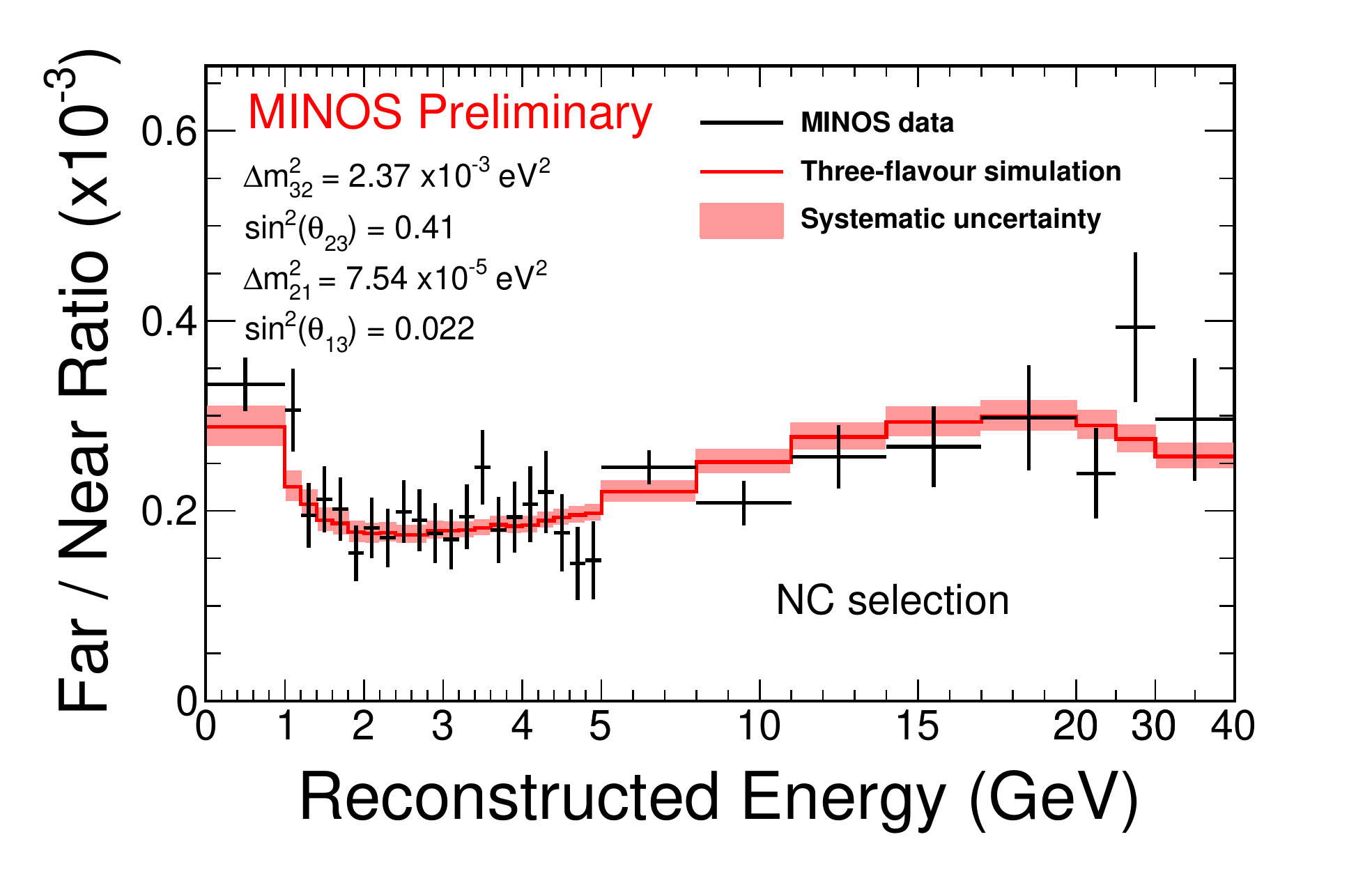}
\caption{Data and Monte Carlo F/N ratios for CC (left) and NC (right) selected events. The red histograms in both panels are the three-flavour expected F/N simulation with systematic error band. The values used for the three-flavour simulation were not fitted to this spectrum but taken from the latest MINOS three-flavour analysis~\protect\cite{ref:threeflavPRL}.}
\label{fig:FNratiso}
\end{figure}

This analysis takes a different approach by fitting the F/N ratio for both CC and NC data sample simultaneously; the F/N ratios for the data and for a three-flavour prediction are shown in figure~\ref{fig:FNratiso}. 

Instead of incorporating systematics as nuisance parameters they are incorporated as uncertainties on the F/N ratio in each bin of energy from $0-40$~GeV using a covariance matrix. The value of the observed F/N ratio is denoted by $\mathbf x$ and the simulated F/N prediction as a function of the oscillation parameters is $\boldsymbol{\mu}$. The likelihood is then computed for values of the oscillation parameters that will minimise the equivalent $\chi^2$ distribution, expressed as:  

\begin{equation}
	\chi^{2} =  \left(\mathbf{x} - \boldsymbol{\mu}\right)^{T}V^{-1}\left(\mathbf{x} - \boldsymbol{\mu}\right)+ \frac{(N_{\text{D}}-N_{\text{MC}})^{2}}{\sigma_{ND}^{2}},
\label{eqn:fit}
\end{equation}   

\noindent where the second term in equation~\ref{eqn:fit} provides a constraint on the absolute neutrino flux at the ND, where $N_{\text{D}}$ and $N_{\text{MC}}$ represent the integrated ND reconstructed neutrino energy spectrum for the $0-40~\text{GeV}$ energy window used in the fit. Flux uncertainties in the neutrino community are known to be difficult to measurement and have large uncertainties associated with them, the error on the ND flux was conservatively set such that $\sigma_{\text{ND}} = 50\%\,\,N_{\text{MC}}$. 

\begin{figure}[!ht]
\centering
\includegraphics[height=0.5\textwidth]{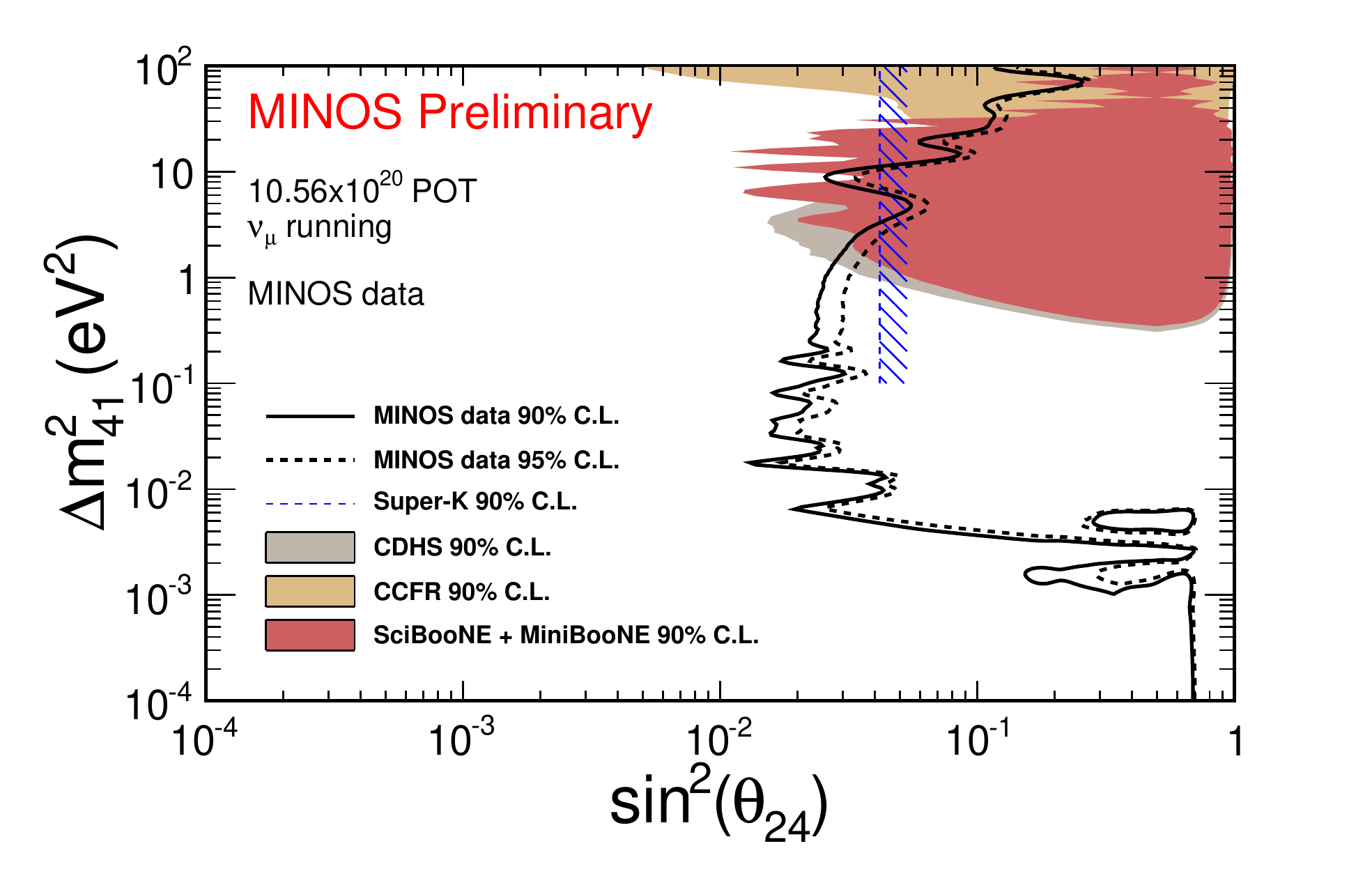}
\caption{Showing the  MINOS 90\% and 95\% C.L. compared to other $\nu_{\mu}$ disappearance experimental measurements. The MINOS limits are produced using the unified procedure of Feldman and Cousins.}
\label{fig:contourMINOS}
\end{figure}

The fit is performed such that a two dimensional likelihood surface is constructed in the plane $\{\Delta m^{2}_{41},\sin^{2}\theta_{24}\}$, over the range $\Delta m^{2}_{41} \in [10^{-4}, 10^{2}]~\text{eV}^{2}$ and $\sin^2\theta_{24} \in [10^{-4}, 1]$. At each point $\Delta m^{2}_{41}$ and $\sin^2\theta_{24}$ are kept fixed with $\Delta m^{2}_{32}$, $\theta_{23}$ and $\theta_{34}$ allowed to vary. All CP violating phases are set to zero as well as $\theta_{14}$; the solar parameters are fixed to global values~\cite{ref:fogli} and $\theta_{13}$ is taken from the weighted average from reactor experiments~\cite{ref:DBth13,ref:Renoth13,ref:choozth13}. 

The results of the 3+1 fit are shown in figure~\ref{fig:contourMINOS}; the 90\% and 95\% confidence limits are displayed, the limits computed have been produced using the Feldman-Cousins unified procedure~\cite{ref:FC}. The contours exclude the region of parameter space to the right of the lines, this analysis sees no statistically significant disagreement with three-flavour model  and sets a strong limit in regions of previously unexplored parameter space. The MINOS result is displayed along with other experimental searches that have used the muon neutrino disappearance channel~\cite{ref:SUPERK,ref:CDHSW,ref:CCFR,ref:SCIBO}. 

\begin{figure}[!ht]
\centering
\includegraphics[height=0.5\textwidth]{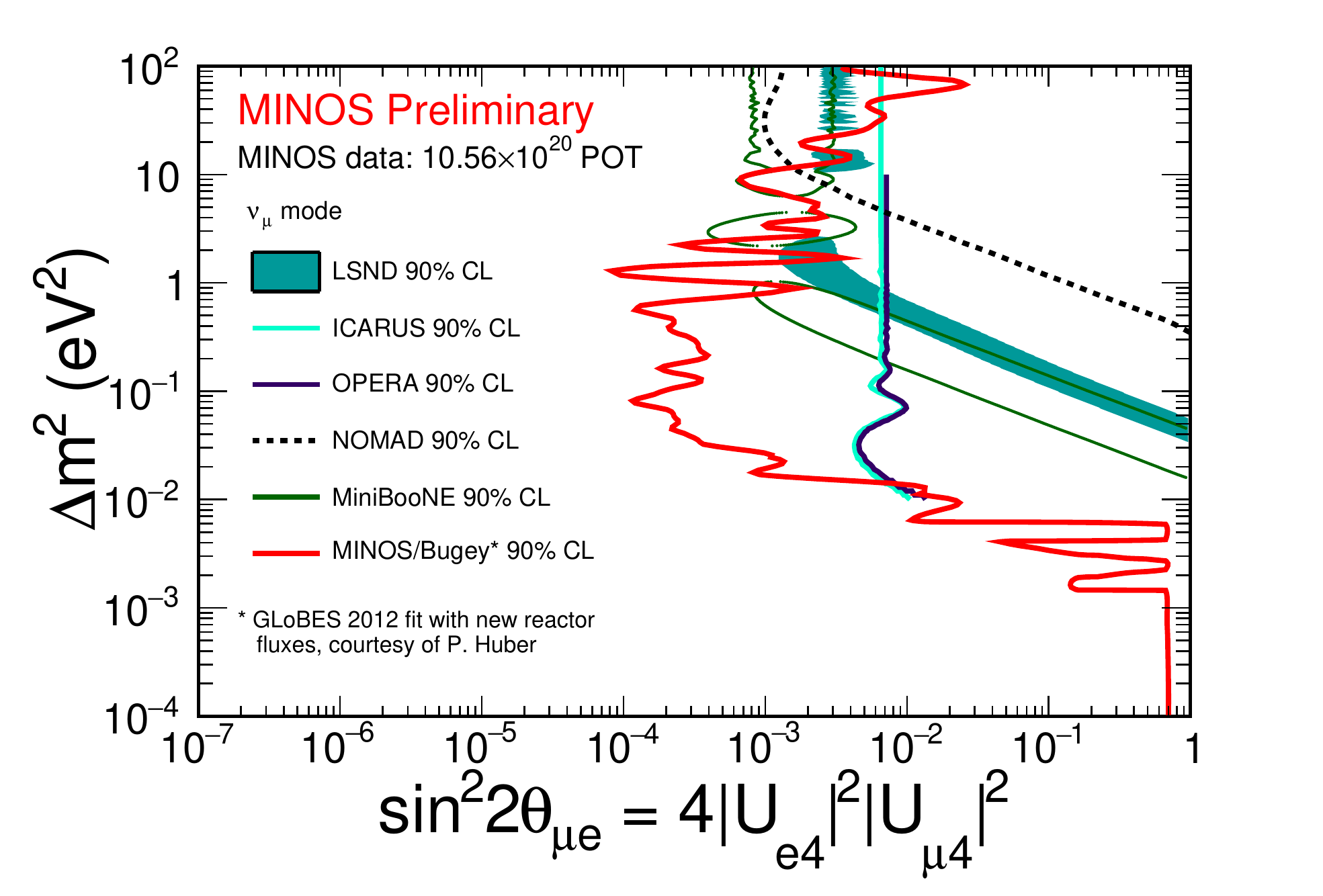}
\caption{MINOS and Bugey combined 90\% confidence level limit on the sterile mixing parameter ${\sin}^22{\theta}\mu e=4{|U{e4}|}^2{|U{\mu4}|}^2$, obtained from the individual disappearance limits of each experiment on the size of ${|{U{\mu4}}|}^2$ and $|{U{e4}}|^2$, respectively. Regions of parameter space to the right of the red contour are excluded at 90\% CL. The MINOS and Bugey result is compared to other electron neutrino appearance experimental results.}
\label{fig:contourCombined}
\end{figure}

MINOS uses the muon disappearance channel giving sensitivity to the sterile mixing angle $\theta_{24}$. Throughout the analysis the assumption that $\theta_{14} = 0$ is taken and therefore by placing limits on the mixing angle $\theta_{24}$ MINOS is placing constraints on the matrix element $|U_{\mu4}|^{2}$. In terms of mixing angles this effective sterile mixing angle can be expressed as $\sin^2 2\theta_{\mu e} = \sin^2 2 \theta_{14} \sin^2 \theta_{24} $. A reactor experiment looking at electron anti-neutrino disappearance from an electron anti-neutrino source will be analogous to the MINOS case although would be sensitive to $\theta_{14}$ and therefore the matrix element $|U_{e4}|^2$. Figure~\ref{fig:contourCombined} shows the 90\% C.L. from the combination between the MINOS and Bugey~\cite{ref:bugey} experiment; during the combination the systematics between both experiments are taken to be uncorrelated. The combined limit is compared to LSND~\cite{ref:LSND} looking at electron anti-neutrino appearance $\overline{\nu}_{\mu} \rightarrow \overline{\nu}_{e}$, and several limits for electron neutrino appearance $\nu_{\mu} \rightarrow \nu_{e}$: MiniBoonE~\cite{ref:MiniBooNE}, OPERA~\cite{ref:OPERAsterile}, ICARUS~\cite{ref:ICARUSsterile}, and NOMAD~\cite{ref:NOMADsterile}.

\section{Discussion and Outlook}

MINOS has performed a 3+1 fit to both the CC and NC F/N ratios and set competitive limits in the parameter space $\{\Delta m^2_{41}, \sin^2\theta_{24}\}$. The results are compatible with the three-flavour model, MINOS combines the result with Bugey allowing for an exclusion of a significant amount of the parameter space suggested flavoured by the LSND and MiniBooNE anomalous results. 

MINOS+~\cite{ref:MINOSPlus} is the continuation of the MINOS detectors but receiving a beam of neutrinos shifted towards higher energies. The beam peak in this higher energy configuration shifts from 3~GeV to 7~GeV allowing MINOS+ to observe around 4,000 $\nu_{\mu}$-CC interactions in the FD each year. MINOS+ has been taking data since September 2013, and with the additional statistics at high energies MINOS+ will be able to significantly extend the reach of its searches for sterile neutrino signatures in the regions of parameter space suggested by LSND and MiniBooNE.

\section*{Acknowledgments}

The work of the MINOS and MINOS+ collaborations is supported by the US
DoE, the UK STFC, the US NSF, the State and University of Minnesota,
the University of Athens in Greece, and Brazil's FAPESP and CNPq. We
are grateful to the Minnesota Department of Natural Resources, the
crew of the Soudan Underground Laboratory, and the personnel of
Fermilab, for their vital contributions.


\end{document}